\documentclass[12pt,preprint]{aastex}
\usepackage{amssymb,latexsym,amsmath,graphics}

\newcommand{\myemail}{yavuz@sabanciuniv.edu}

\shorttitle{Fall-Back Disks \& AXPs}
\shortauthors{EK\c{S}\.{I} \& Alpar  }

\begin{document}

\title{Can Thin Disks Produce Anomalous X-Ray Pulsars?}

\author{K. YAVUZ EK\c{S}\.{I}\altaffilmark{1,2} and M. AL\.{I} ALPAR\altaffilmark{2}}

\email{ \myemail, alpar@sabanciuniv.edu}

\altaffiltext{1}{Bo\u{g}azi\c{c}i University, 34342, Bebek, \.{I}stanbul, Turkey}
\altaffiltext{2}{Sabanc\i\ University, Orhanl\i--Tuzla, \.{I}stanbul 34956, Turkey}

\begin{abstract}
We investigate whether young neutron stars with fall-back disks can produce
Anomalous X-Ray Pulsars (AXPs) within timescales indicated by the ages of
associated supernova remnants. The system passes through a propeller stage
before emerging as an AXP or a radio pulsar. The evolution of the disk is
described by a diffusion equation which has self-similar solutions with
either angular momentum or total mass of the disk conserved. We associate
these two types of solutions with accretor and propeller regimes,
respectively. Our numerical calculations of thin disk models with changing
inner radius take into account the super-critical accretion at the early
stages, and electron scattering and bound-free opacities with rich metal
content. Our results show that, assuming a fraction of the mass inflow is
accreted onto the neutron star, the fall-back disk scenario can produce AXPs
for acceptable parameters.
\end{abstract}


\section{INTRODUCTION}

Anomalous X-Ray Pulsars (AXPs) \citep{mereghetti} are distinct
from accretion powered X-ray pulsars in several aspects: (1) No
binary companion or orbital Doppler modulations have been
observed. (2) AXP rotation periods are clustered between 5-12 s
while the periods of conventional X-ray pulsars span a much wider
range ($P\sim 0.069-10^{4}$ s). (3) They have large spin-down
rates \citep{MS95} with a time-scale of $\tau =P/2\dot{P}\sim
10^{3}-10^{5}$ years, (4) Three of them have been associated with
supernova remnants (SNRs) indicating that they are young
($t_{SNR}\lesssim 5\times 10^{4}$ yrs.) \citep{gaensler, tagieva},
(5) Luminosities of $L_{x}\sim 10^{35}-10^{36}$ ergs s$^{-1}$ are
well in excess of the spin-down power, (6) AXP spectra are soft
compared to typical X-ray pulsars, with power law indices $\Gamma
\gtrsim 2$. AXPs share all these features with Soft Gamma Ray
Repeaters (SGRs) \citep{hurley,kouve98,kouve99}. The observation
of bursts from AXPs \citep{AXPburst1, AXPburst2},  suggests a
strong connection between AXPs and SGRs.

In magnetar models \citep{duncan92,thompson95,thompson96}
bursts are triggered and powered by enormous magnetic fields, $B\sim 10^{15}$
Gauss, and the energy source of the X-Ray emission is the decay of this
magnetic field \citep{thompson95,CGP}. The star spins down by
magnetic dipole radiation. While the magnetar model is quite successful in
modelling the SGR and AXP bursts, and spin-down rates, it can not explain
the period clustering \citep{PM02} except in one set of field decay models
under special conditions \citep{CGP}.

Accretion models for AXPs started with the work of
\citet{vanParadijs}. The current models invoke neutron stars with
$B\sim 10^{12}$ G and explain the period clustering in terms of
asymptotic evolution of the neutron star rotation rate towards
equilibrium with a fall-back disk (Chatterjee, Hernquist \&
Narayan 2000, hereafter CHN; Alpar 2001; Marsden, Lingenfelter \&
Rothschild  2001). These models require no binary companion. AXP
ages are indicated by the ages of the supernova remnants
associated with some AXPs. The possibility that some material in a
supernova explosion might fall back and accrete onto the new-born
neutron star has been explored by several authors
\citep{colgate,SP71,RS73}. Fall-back accretion disks have been
invoked to address a diversity of astrophysical problems
\citep{MD81,MD83,michel88,lin91}. \citet{mineshige97} have shown,
by using smoothed particle hydrodynamics \citep{SPH}, that an
accretion disk is formed around a new-born compact object if the
progenitor had been rotating before the explosion. The total mass
of the fall-back gas has been estimated to be $0.05M_{\sun }$ by
\citet {hashimoto}, $0.1M_{\sun }$ by \citet{BB94} and
$0.15M_{\sun }$ by \citet{cheva89} (see also \citet{lin91}).

The angular momentum carried by the neutron star and by the
ambient material must play an important role in determining the
subsequent evolution. Motivated by the recognition of angular
momentum as an important initial parameter, \citet{alpar01}
proposed a classification of young neutron stars in terms of the
absence or presence and properties of a fall-back disk. According
to this model AXPs, SGRs and Dim Thermal Neutron Stars (DTNs) have
similar periods because they are in an asymptotic spin-down phase
in interaction with a fall-back disk. The different classes
represent alternative pathways of neutron stars. Radio pulsars
have no disks or encounter very low mass inflow rates while Radio
Quiet Neutron Stars (RQNS) have such high mass inflow rates that
their pulse periods are obscured by the dense medium around. AXPs
and SGRs evolve through a propeller stage
\citep{shv70,PR72,IS75,fabian75, ihsan} in which the rapid
rotation of the neutron star prevents the inflowing matter from
accreting onto the surface of the star and the star is spun-down
until its rotation is slow enough that accretion can commence.

CHN pioneered a specific evolutionary model to produce AXPs with a
fall-back disk. They employed the available time-dependent viscous
thin disk models. There are upper limits and observations in the
infrared and optical that have been evaluated within the CHN thin
disk model to constrain and argue against the fall-back disk
models in general. Two points should be underlined here. First,
the recent detections in the infrared (and optical) supports the
existence of a disk \citep{israel03}. The magnetar models do not
have a prediction for magnetospheric production of infrared
photons, and the detections in IR are well in excess of the
extrapolation of a neutron star surface black-body fit to the
X-ray band. The second point is that if a disk is present, it is
unlikely that it will be a thin disk in a stable state, as the
sources are expected to spend significant parts of their evolution
in a propeller phase in which mass that can not be accreted to the
neutron star will remain bound around the star and disk. In
addition, there will be irradiation effects on the disk. The thin
disk models are probably not the realistic models to compare with
the data. Nevertheless, as the only available working models, they
warrant a careful study to decide whether interaction with a
fall-back disk can lead to AXP periods and luminosities at the
ages indicated by SNR-AXP associations. On the other hand optical
pulsations with a large pulse fraction observed from
\objectname{4U 0142+614} \citep{KM02} can not be explained by
emission or reprocessing from a disk. This observation, in
addition to its successful application to model SGR and AXP
bursts, clearly favors magnetar models, while the period
clustering has a natural explanation in disk models. It may be the
case that the AXPs (which show bursts like SGRs) do have surface
magnetic fields in the higher multipoles and a dipole field in the
$10^{12}$ G range interacting with a fall-back disk to provide the
rotational equilibrium that leads to period clustering. Such a
hybrid model in which AXPs and SGRs would be the rare objects that
correspond to the high magnetic field \emph{and} high angular
momentum (fall-back disk) corner of the parameter space of initial
conditions, makes the actual possibilities of disk models worth
exploring.

An investigation of the CHN model was carried out by \citet{FW} who
argued that with the bound-free opacities that are relevant, the evolution
would not lead to AXPs. This work was carried
out with a disk solution with inner boundary condition relevant for
a disk losing mass by accretion whereas the system throughout
its evolution is in the propeller regime.

We employ different thin-disk solutions for propeller and for accretor
phases of the evolution, use electron scattering and bound-free (mostly the
latter) opacities as appropriate and include a range of initial disk masses.
In the next section we present the model equations, the special analytical
solutions, and their relation to the numerical calculations for a finite
disk. Results are presented in section 3, and discussed in the final section.

\section{THIN DISK MODELS}

The evolution of a viscous thin disk is governed by a non-linear diffusion
equation \citep{pringle81} which accepts three self-similar solutions
\citep{pringle74} extending down to $r=0$, if the disk opacity has a power-law
dependence on density and temperature. One of these solutions is a simple
power-law both in spatial and temporal variables and is irrelevant for
fall-back disks which have freely expanding outer boundaries. The other two
solutions can be associated with accretion and propeller regimes. In the
solution which we associate with the accretion regime \citep{CLG} the total
angular momentum of the disk is constant ($\dot{J}_{d}=0$) and mass of the
disk decreases in time such that $\dot{M}_{d}\propto t^{-\alpha }$. This is
the solution employed by CHN for all phases of AXP evolution.
In the other
solution which we associate with the propeller regime, the mass of the disk
is constant ($\dot{M}_{d}=0$) and angular momentum of the disk increases in
time by the viscous torque at the inner radius such that $\dot{J}_{d}\propto
t^{-\beta }$. \citet{pringle81} mentioned that  ``such a
solution might represent a disc around a magnetized star which is rotating
sufficiently rapidly that its angular velocity exceeds the Keplerian
velocity at the magnetosphere'' - i.e. a propeller. A real disk
starts at an inner radius $R_{m}$ determined by the magnetic moment $\mu $
and rotation rate $\Omega _{\ast }$ of the neutron star, and mass flow rate
$\dot{M}(r)$. For both types of analytical solutions extending to $r=0$, the
application to a real disk with finite inner radius $R_{m}$ requires a
numerical treatment, which shows that the accretor solution \citep{CLG} and
the propeller solution \citep{pringle91} are good representations of the
numerical solutions, the initial conditions being `forgotten' after a brief
transition period. One dimensional thin disk solutions necessarily
suppress the two dimensional mass flow in the $r-z$ plane of a realistic
disk model. In the case of accretor type analytical solutions, the mass flow
$\dot{M}(r)$ is inward in the inner disk. In the propeller type analytical
solutions $\dot{M}(r)$ is outward, with the mass loss rate of the disk
$\dot{M}_{d}\equiv \dot{M}(r=0)=0$. (\citet{pringle91} numerically shows  that there
are also propeller type solutions ---with non-vanishing torque at the inner boundary--- in which
mass flow is inwards in the inner parts of the disk.) In the real propeller, constant disk mass would
be sustained with mass inflow and outflow; and small fractions of the mass
flow can accrete to the neutron star or escape the system. For opacities of
the form $\kappa =\kappa _{0}\rho ^{a}T^{b}$, thin disk equations \citep{FKR92}
 can be solved to yield the power law indices of the time dependence
in the two analytical solutions:
\begin{equation}
\dot{M}_{d}\varpropto t^{-\alpha },\qquad \alpha =\frac{18a-4b+38}{17a-2b+32}
\label{alpha}
\end{equation}
for the accretor type solution \citep{CLG,FW}; and
\begin{equation}
\dot{J}_{d}\varpropto t^{-\beta },\qquad \beta =\frac{14a+22}{15a-2b+28}.
\label{beta}
\end{equation}
for the solution we associate with propeller regime \citep{pringle91}. For electron scattering
dominated disks ($a=b=0$) $\alpha =19/16$ and $\beta =11/14$, and for
bound-free opacity dominated disks ($a=1$, $b=-7/2$) $\alpha =5/4$ and $\beta =18/25$. CHN used
$\alpha=7/6$ for analytical convenience.

The total torque is the sum of accretion torque
$\dot{J}_{a}=r^2\Omega \dot{M}$ and the viscous torque
$\dot{J}_{\nu}=2\pi r^2 W$ where $W$ is the vertically averaged
viscous stress. The accretor solution is not precisely the
time-dependent version of the \citet{SS73} solutions with the same
inner boundary condition. The latter solutions assume vanishing
\emph{viscous torque}  whereas the time-dependent solution implies
vanishing \emph{total torque} at the inner boundary.

The inner boundary of the disk, $R_{m}$, is where the ram pressure
$P_{ram}=\rho v^{2}$ of the disk is balanced by the magnetic pressure
$P_{mag}=B^{2}/8\pi $. Assuming a dipole field for the neutron star, the
magnetic pressure can be written as \citep{lipunov}:
\begin{equation}
P_{mag}=
\begin{cases}
\frac{\mu ^{2}}{8\pi r^{6}} & \text{if $r\leq R_{L}$;} \\
\frac{\mu ^{2}}{8\pi R_{L}^{4}r^{2}} & \text{if $r>R_{L}$.}%
\end{cases}
\label{Pmag}
\end{equation}
where $R_{L}=c/\Omega _{\ast }$ is the light cylinder radius and $\mu $ is
the dipole magnetic moment. We write the ram pressure as%
\begin{equation}
P_{ram}=
\begin{cases}
\frac{\dot{M}}{4\pi r^{2}}\sqrt{2}\Omega _{K}r & \text{if $r\leq R_{co}$;}
\\
\frac{\dot{M}}{4\pi r^{2}}\sqrt{2}\Omega _{\ast }r & \text{if $r>R_{co}$.}%
\end{cases}
\label{Pram}
\end{equation}
(see \citet{romanova}) where $R_{co}=\left( GM/\Omega _{\ast
}^{2}\right) ^{1/3}$ is the corotation radius and $\dot{M}$ is the
mass flow rate. Here $\Omega _{K}$ denotes the Keplerian rotation
rate at the inner edge of the disk. The two expressions we adopt
for $P_{ram}$ are asymptotic forms for $\omega _{\ast }\ll 1$ and
$\omega _{\ast }\gg 1$ where $\omega _{\ast }=\Omega _{\ast
}/\Omega _{K}(R_{m})$ is the fastness parameter.

When the flow rate at $R_{m}$ is greater than the critical rate
$\dot{M}_{Edd}=4\pi m_{p}cR_{m}/\sigma _{T}$, we follow \citet{SS73} and assume that
the disk inside the ``spherization'' radius
$R_{s}=\sigma _{T}\dot{M}/4\pi m_{p}c$ regulates itself as
\begin{equation}
\dot{M}\left( r\right) =\frac{r}{R_{s}}\dot{M}_{Edd}\quad \text{if $r\leq
R_{s}$;}  \label{regulate}
\end{equation}
with the result that the total luminosity of the disk does not exceed the
Eddington limit considerably \citep{lipunov,lipunova99}. These
considerations are relevant for the earliest stages of the mass flow
typically extending over $\sim 20$ years in our calculations and do not
effect the later period evolution of the star significantly. However, without such an incorporation
of radiation pressure, the initial inner radius of the disk is obtained to be smaller than the radius
of the neutron star, which is unphysical, and the system passes through an accretor stage. With the
implementation of radiation pressure effects, the initial inner radius of the disk is greater than the
corotation radius and the system is in a supercritical propeller stage.
In the numerical calculations, the inner radius
$R_{m}=R_{m}(\dot{M},\Omega _{\ast})$ is determined self-consistently at each time step
using equations (\ref{Pmag}), (\ref{Pram}) and (\ref{regulate}).

The accretor solution with the boundary condition $\dot{J}_{d}=0$
had been applied for fall-back disks by
\citet{mineshige93,MPH01,PHN00}, and CHN.  Starting from the early
phases, throughout most or all of its evolution into the
asymptotic ``tracking'' AXP phase, the disk is in the propeller
regime. We employ the relevant propeller type disk solution. This
follows the power-law solution for $\dot{J}_{d}$ given in Eq.
(\ref{beta}). In these propeller solutions at the inner boundary
$\dot{M}(R_{m})=0$ while in the accretion regime
$\dot{M}(R_{m})=\dot{M}_{d}$. We shall define a representative
value $\dot{M}$ of the mass flow rate in the inner disk from the
propeller torque $\dot{J}_{d}$ translating the torque
$\dot{J}_{d}\propto t^{-\beta }$ to a flow rate such that
\begin{equation}
\dot{J}_{d}\propto \sqrt{GMR_{m}}\dot{M}.
\label{translate}
\end{equation}
This yields a representative mass flow rate $\dot{M}\propto
t^{-\alpha _{P}}$, defining the effective index $\alpha _{P}$. We
emphasize that this definition is an estimate of the mass flow
rates within the disk while the disk mass remains constant, with
no mass loss from boundary. By employing the solution with a
constant disk mass, we implicitly assume that the inflowing matter
flung away by the magnetosphere of the neutron star returns back
to the disk somewhere away from the inner boundary of the disk.

The critical fastness parameter for transition from propeller to accretor regime is assumed to be
$\omega_{cr}\gtrsim 1$ and we take it to be 1 in all numerical experiments. The fastness parameter at which the torque vanishes
is the equilibrium fastness
parameter $\omega_{eq}$ and is inferred to be somewhat smaller than $\omega_{cr}$ as there are
systems, like AXPs in our model, which are spinning down while accreting. We have observed numerically
that the results do not depend
sensitively on the value of $\omega_{eq}$ and we take it to be $0.7$.
The range $\omega_{eq}<\omega _{\ast }<\omega_{cr}$
is not well interpreted with either the propeller or the accretor type of  disk solutions, so
when the fastness parameter $\omega _{\ast }$ has decayed to  $\omega _{cr}$,
we try two alternatives:
(i) switching to the accretor solution (arguing $\dot{M} \neq 0$), (ii) continuing with the propeller
solution (arguing $\dot{J} \neq 0$). In both cases we assume that a small fraction
$\eta=\dot{M}_{accreted}/ \dot{M}$ of the mass flow at the magnetic radius can reach the surface of the neutron
star in order that $L=GM\eta \dot{M}/R_{\ast }$ gives the AXP luminosities.

\citet{FW} showed that the disk is electron scattering dominated only in the initial stages when
$\dot{M}_{17}( t) >50[( 1+X) /Z]^{-7/10}B_{12}^{3/5}$,
while later on, at lower mass flow rates, bound-free opacity
dominates. Here $\dot{M}_{17}=\dot{M}/10^{17}$g s$^{-1}$, $B_{12}=B/10^{12}$ Gauss,
$X$ is the hydrogen mass fraction and $Z$ is the heavy element mass
fraction. Assuming the disk to be rich in heavy elements, we choose $X=0.1$
and $Z=0.9$. The flow rate in the propeller solution declines with a softer
power-law index than it does in the accretion solution because in the former case
matter cannot accrete and the decrease in the flow rate is only by the
viscous spreading of the disk.

We incorporate the torque model used by \citet{alpar01} in the form
\begin{equation}
\dot{J}_{\ast } =
\begin{cases}
\sqrt{GMR_{m}}\ \dot{M}\ \left( 1-\omega _{\ast }/\omega _{eq }\right)
& \text{ for $\omega_{\ast }<\omega_{cr}$;}
\\
\dot{J}_{d}\left( 1-\omega _{\ast }/\omega _{eq }\right)
& \text{ for $\omega_{\ast}>\omega_{cr}$.}
\end{cases}
 \label{torque}
\end{equation}
As noted by \citet{FW} and \citet{li02}, this is a very efficient
propeller torque model compared to some other available propeller
torques obtained through simple energy or angular momentum
arguments. This form of the torque is supported by the detailed
numerical simulations of the magnetic boundary layers investigated
by \citet{daumerie}. Recently \citet{ihsan} revived the propeller
model of \citet{DP81} which incorporates a still more efficient
torque $\dot{J} \propto -\omega_{\ast}^2$ for the sub-sonic
propeller, and a torque $\dot{J} \propto -\omega_{\ast}$, which
agrees with our model in the large $\omega_{\ast}$ limit, for the
super-sonic propeller regime. These models do not entail
asymptotic approach to rotational equilibrium, and are therefore
not relevant for addressing the period clustering of AXPs. The
recent numerical work of \citet{romanova} also indicate an
efficient propeller torque ($\dot{J}\propto
-\Omega_{\ast}^{4/3}$). For $\omega_{\ast} \gg 1$ matter can be
flung away with the tangential velocity of the field lines at
$R_m$ which is greater than the escape velocity so that some mass
loss will take place in disagreement with the propeller type
solution we employ. This means that the remaining mass of the
disk, when the system reaches $\omega_{\ast}=\omega_{cr}$ and
accretion onto the neutron star is allowed, is actually lower than
what we find by assuming the mass of the disk is conserved. In
this case the factor $\eta=\dot{M}_{accreted}/ \dot{M}$ need not
be very low but can attain a reasonable value.

The initial mass of the disk can be found by integrating the mass
flow rate after accretion starts. This yields
$M_{d}=t_{tr}\dot{M}_{tr}/(\alpha-1)$ where  $t_{tr}$ is the
transition time from propeller to the accretor regime and
$\dot{M}_{tr}$ is the inflow rate at  $t=t_{tr}$, and
$\alpha=5/4$. By fitting to the numerical solutions we have
determined, for an initial mass flow rate of $5\times 10^{27}$ g
s$^{-1}$, that $\dot{M}_{tr}=2\times 10^{16}\times
\mu_{30}^{1.96}$ g s$^{-1}$ and $t_{tr}=10^5\times \mu_{30}^{-2}$
years which means that $M_d \cong 1.2 \times 10^{-4}M_{\sun}$.
This nominal value of the initial disk mass  is 50 times smaller
than the nominal value incorporated in the CHN model. From our
fits we find that the initial inflow rate and the initial mass of
the disk are related through
$M_d/(10^{-4}M_{\sun})=0.2074\dot{M}_0/(10^{27}\text{g
s$^{-1}$})+0.1812$ for our parameter range.

\section{RESULTS}

We have followed the evolution of the star-disk system for
$10^{5}$ years, solving the torque equation (\ref{torque})
numerically with the adaptive step-size Runge-Kutta method
\citep{NR}. We assumed the radius and mass of the neutron star to
be $R_{\ast}=10^6$ cm and $M_{\ast}=1.4M_{\sun}$, respectively. In
all numerical experiments we have fixed the initial period
$P_{0}=15$ ms, and considered dipole magnetic moments in the range
$\mu _{30}=1-10$. Electron scattering opacity prevails in the disk
at the high initial flow rates. The system is initially in the
super-critical propeller stage (i.e. radiation pressure is
dominant). To follow one example, for an initial mass flow rate
$5\times 10^{27}$ g s$^{-1}$ which corresponds to an initial disk
mass of $1.2\sim 10^{-4}M_{\sun}$ and the inner radius is at
$R_{m}=31R_{\ast }\mu_{30}^{1/3}\left( P_{0}/15ms\right) ^{1/6}$.
The dynamical time-scale $T_{d}\sim \Omega _{K}^{-1}$ at this
radius is $1.23\times 10^{-2}\mu _{30}^{1/6}(P_{0}/15ms) ^{1/12}$
s. For $\mu _{30}=5$ the supercritical propeller stage lasts for
$\simeq 15$ years. In this stage the effective power law index,
$\alpha_p$ is equal to $\beta=11/14$, because the disk inner
radius does not depend on $\dot{M}_0$ in the super-critical
regime, so that $\dot{M}=10^{27}( t/T_{d}) ^{-11/14}$ (see Eq.
(\ref{translate})). After $\simeq 15$ years radiation pressure
ceases to be important. Electron scattering continues as the
dominant opacity until  $t \simeq 75$ years; in this stage the
effective power law index is  $55/63$. After $t \simeq 75$ years,
the propeller regime continues with bound-free opacity dominating
in the disk; the effective power law index is now $\alpha_p =4/5$
and remains so as long as the propeller regime continues. At
around $10^{4}$ years the fastness parameter drops below $\omega
_{cr}$ and the system enters the
 ``tracking phase'' CHN. It is
during this period that we observe the system as an AXP.

\begin{figure}
\plottwo{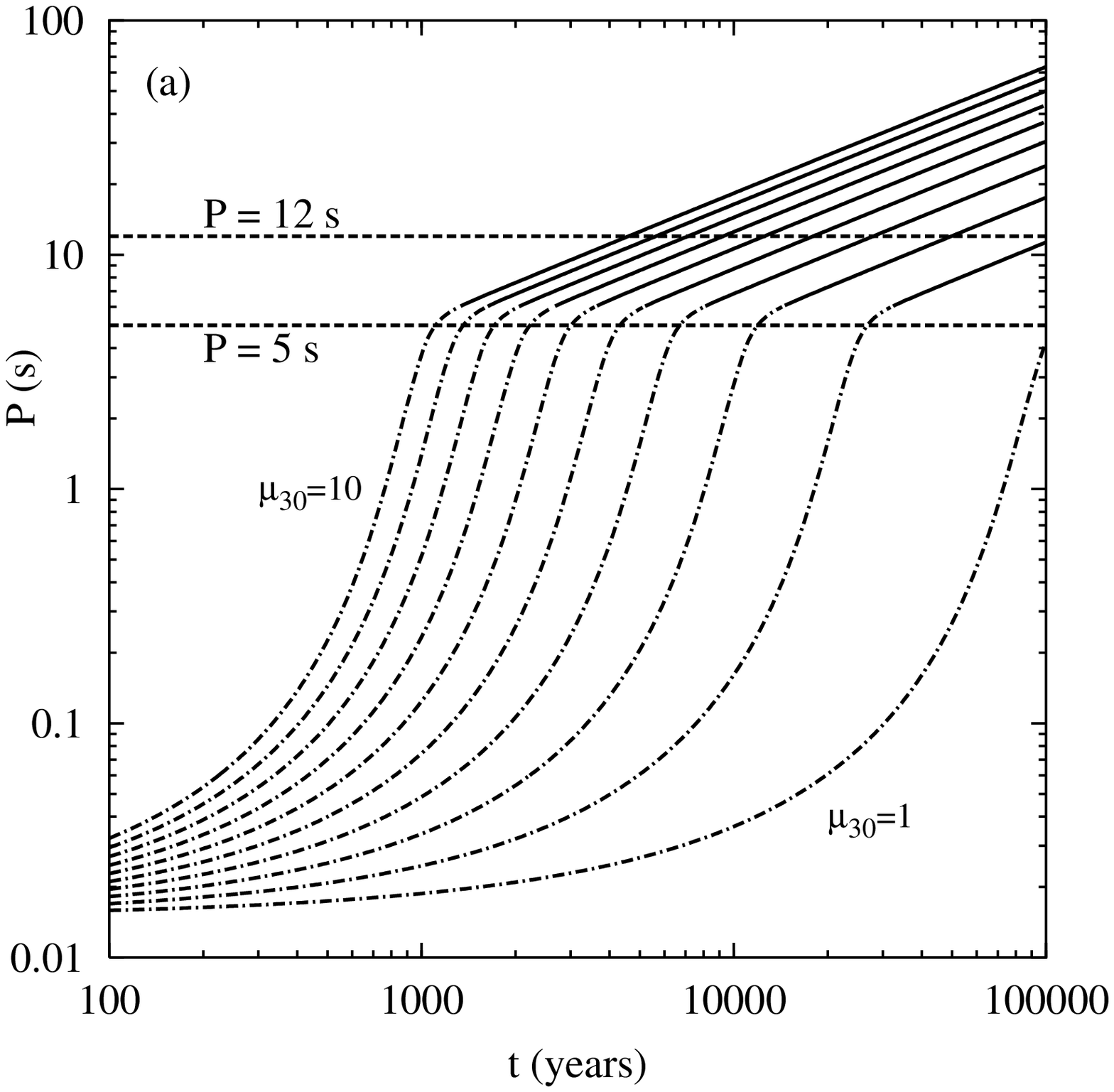}{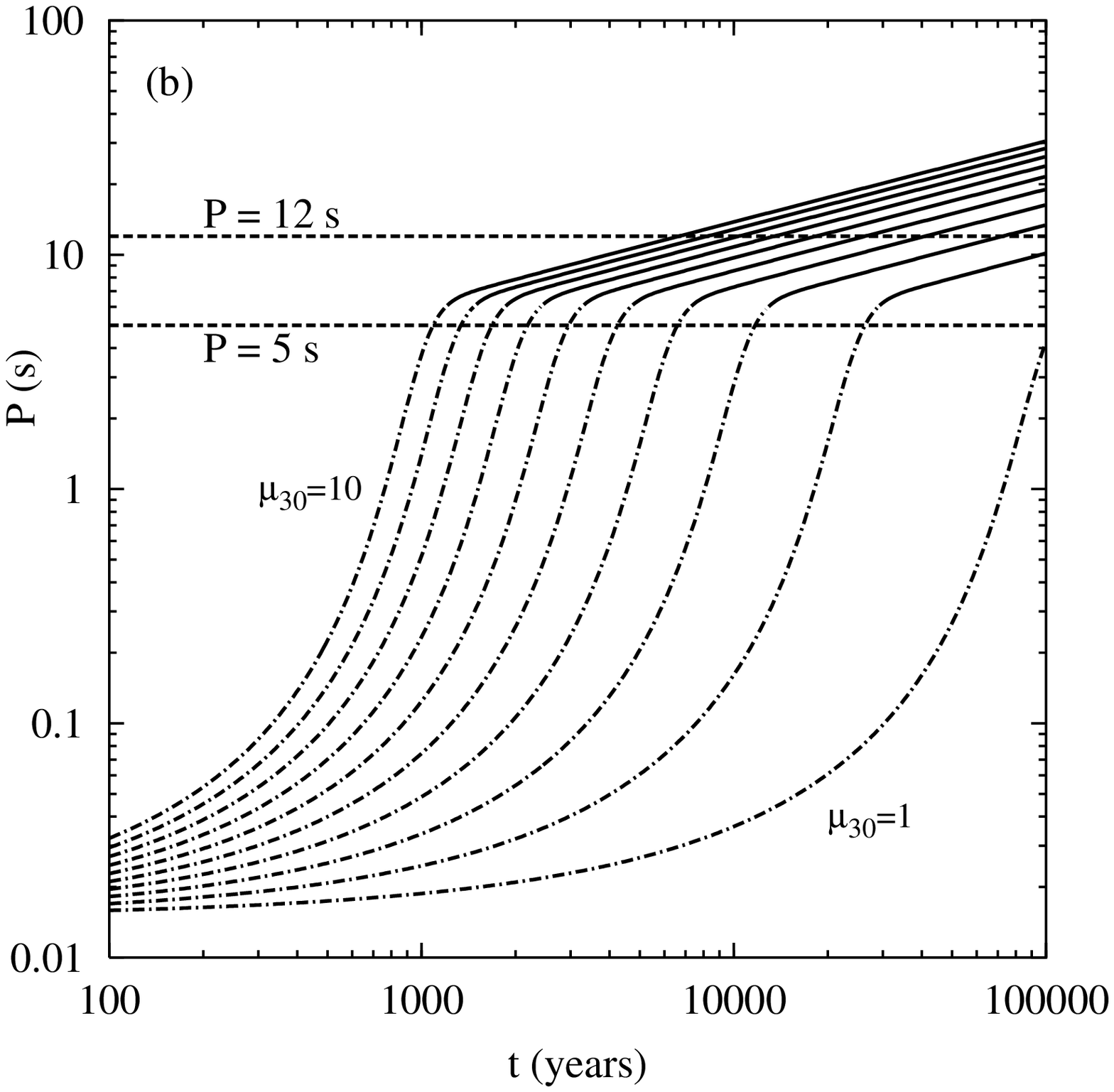} \caption{Period evolution of a neutron
star with mass $M=1.4M_{\sun}$, initial period $P_0=0.015$ s and
initial disk mass  $M_0=1.2\times 10^{-4}M_{\sun}$. The values of
magnetic dipole moments $\mu_{30}$ are 1, 2, \ldots, 10, with
$\mu_{30}=1$ for the lowermost curve and $\mu_{30}=10$ for the
uppermost. The \emph{dashed and dotted lines} show the rapid
spindown phase while the \emph{solid line} stand for the
``tracking'' stage. The \emph{dashed horizontal lines} show the
period range of AXPs and SGRs. Neutron stars with higher magnetic
dipole moments reach the tracking phase earlier. The panels
correspond to the two options we used for
$\omega_{\ast}<\omega_{cr}$.  In both cases $\omega_{cr}=1$ and
$\omega_{eq}=0.7$. In case (a) the disk evolves with the accretor
solution for $\omega_{\ast}<\omega_{cr}$, and in case (b) it
continues to evolve with the propeller solution. \label{periods}}
\end{figure}

\begin{figure}
\plottwo{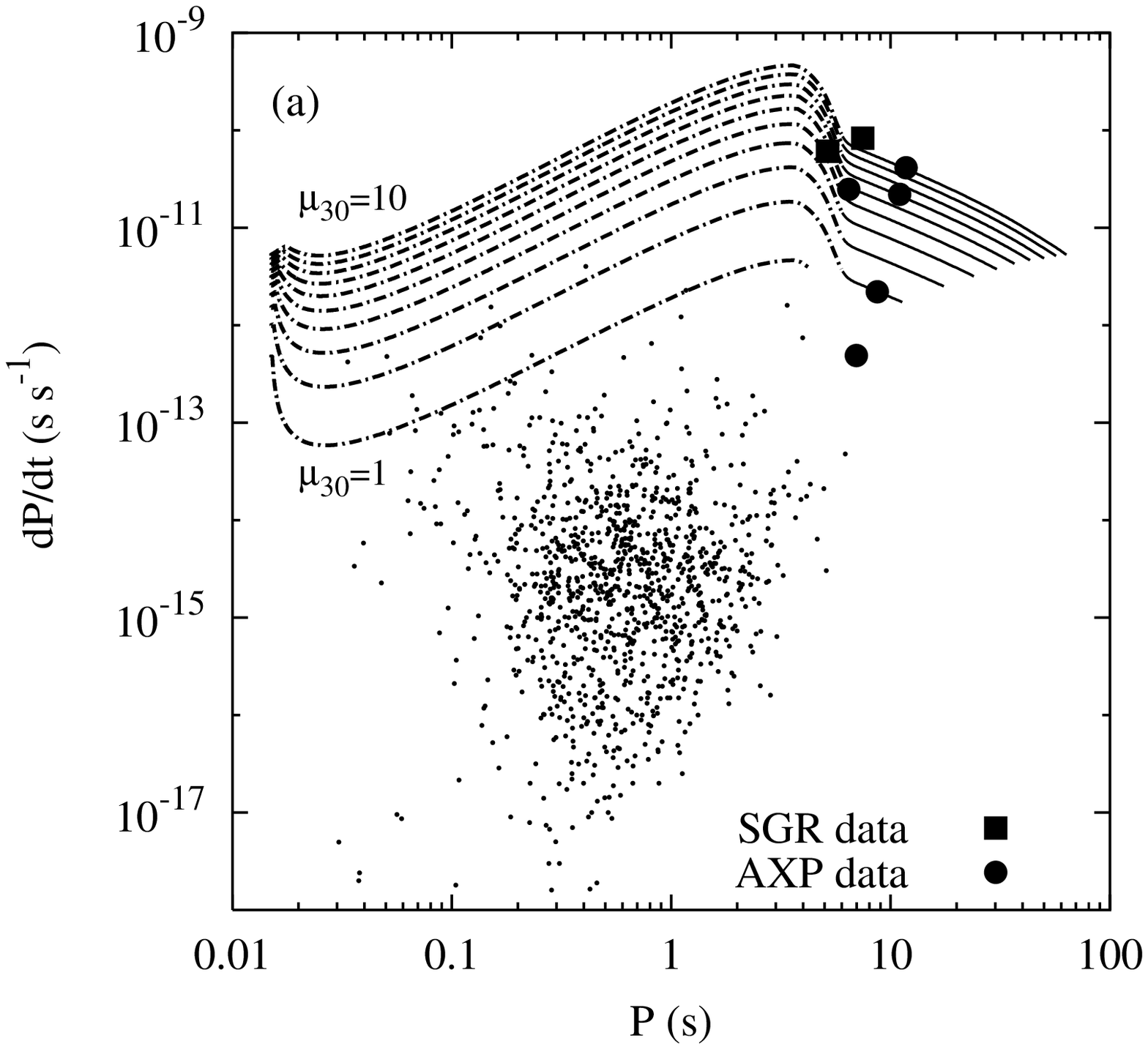}{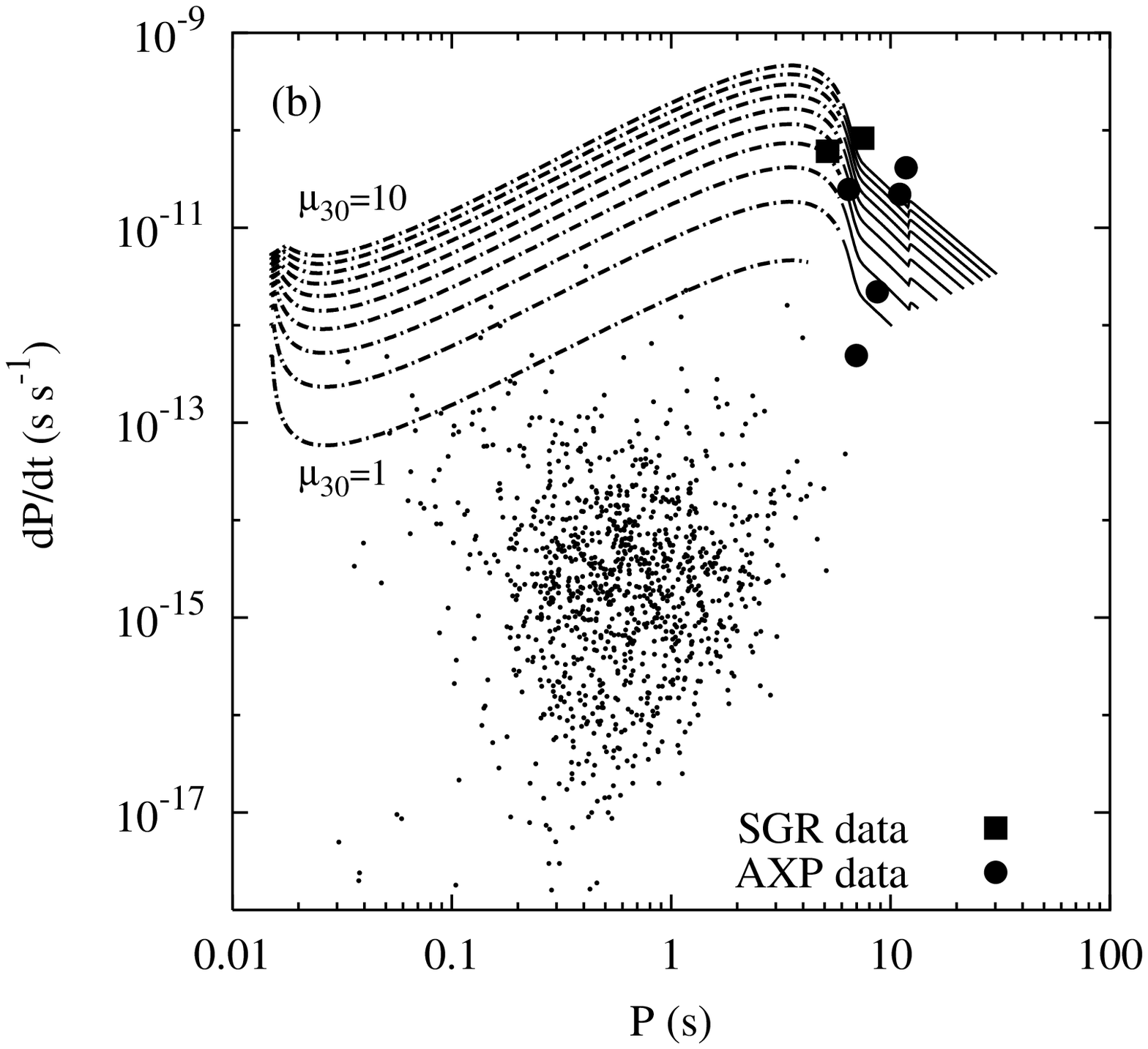} \caption{Evolutionary tracks in the
$P-\dot{P}$ diagram with the same parameters as in
Figure--\ref{periods}. The two panels again correspond to the two
cases in which (a) the disk switches to the accretor solution, and
(b) it proceeds with the propeller solution. The \emph{dashed and
dotted lines} show the rapid spindown phase while the \emph{solid
line} stand for the ``tracking'' stage. The \emph{dots} denote the
radio pulsars. AXPs are shown by \emph{filled circles} and SGRs by
\emph{squares}. With different $\mu_{30}$ and/or  initial mass
values of the disk, the model can cover all AXPs and SGRs.
\label{PPdots}}
\end{figure}

In Figure \ref{periods} we give the period evolution of a neutron
star for these parameters. These are the typical values leading to
AXPs and SGRs. Panel (a) corresponds to the case in which we
switched to the accretor type of solution  when the fastness
parameter decayed below $\omega_{cr}$ and panel (b) correspond to
the case in which we continued with the propeller solution. We see
that the latter case yields a somewhat more gradual period
evolution because the equilibrium period increases less rapidly in
this case. Figure \ref{PPdots} shows the evolutionary tracks in
the $P-\dot{P}$ diagram. Again the two panels (a) and (b)
correspond to the two cases as described for Figure \ref{periods}.
We see that the fall-back disk model can account for the position
of AXPs and SGRs in the $P-\dot{P}$ diagram. In both cases the
observed AXP luminosities require that the mass accretion rate
onto the neutron star, $\dot{M}_{\ast}$, is only a fraction
$\eta=\dot{M}_{\ast}/\dot{M}_d \sim 10^{-2}$ of the mass accretion
rate in the disk. Whether the accreted fraction can be this low
when the system is asymptotically close to equilibrium is an open
question.  The Soft X-Ray Transient Aql X-1 may have exhibited a
comparably small accreted fraction of its mass transfer rate when
it went through a reduction in luminosity by $10^{-2}-10^{-3}$,
interpreted as a transition to the propeller state \citep{campana,
zhang, cui}. Alternatively, the small value of $\eta$ implied by
the present model calculations may reflect our assumption that the
mass flow of the disk remains constant throughout the propeller
phase:  In reality some matter should have been expelled from the
system which would lead to lower mass flow rates in the present
day disk and hence the actual accreted fraction $\eta$, to meet
the observed luminosities, could be higher. In any case models for
an estimation of $\eta$ on the basis of accretion or propeller
dynamics are not available at present (see \citet{romanova} for a
numerical approach). The result $\eta \sim 10^{-2}$ is an
empirical constraint on fall-back disk models if they are to
explain the observed luminosities of AXPs. Any independent
evidence on the value of $\eta$ in these or similar sources will
provide a test of the presently employed fall-back disk model.

\placefigure{periods}
\placefigure{PPdots}

Our results do not change qualitatively for initial disk masses
$0.8-2.25 \times 10^{-4}M_{\sun}$. The transition to the accretion
stage becomes earlier for greater initial disk masses. For $2.25
\times 10^{-4}M_{\sun}$ and $\mu_{30} = 4$, the transition to the
accretion regime occurs at 3000 years and for higher initial disk
masses we obtain mass inflow rates as high as $10^{19}$ g s$^{-1}$
which give $\eta \sim 10^{-3}$ to obtain present day luminosities
of AXPs. For $M_d<0.8 \times 10^{-4}M_{\sun}$ the magnetic radius
exceeds the light cylinder radius $R_L$ at an early stage and the
source becomes a radio pulsar.
 The fact that the model gives radio pulsars at initial disk masses just below the range leading to AXPs and SGRs
 supports a continuity between  AXPs, SGRs and radio pulsars, and suggests that radio pulsars might also have disks around them.
We shall explore in a separate paper the possibilities of this continuity in initial conditions for different classes of neutron stars,
in particular evolutionary tracks leading to AXPs, SGRs \emph{and} radio pulsars in the same region of the  $P-\dot{P}$ diagram as the
AXPs \citep{mclaughlin}.

Thin disk models yield AXPs only if an efficient torque model is
used as noted by \citet{FW} and \citet{li02}. Our simulations with
other torque models \citep{fabian75,GL78} show that these do not
produce AXPs, verifying the results of \citet{FW} and
\citet{li02}. However, as noted in the previous section, recent
analytical \citep{ihsan} and numerical \citep{daumerie, romanova}
studies lend support to an efficient torque model.

The question naturally arises as to why the disk evolution does
not continue, to give sources with $P_{\ast}>12$ s; what is the
justification for assuming that disk evolution stops at a finite
age ($\sim 10^5$ yrs in the present work)? The formal power-law
solutions predict disk masses/torques decaying to infinitely small
values as $t \rightarrow \infty $. In reality at low density and
temperature the disk undergoes a transition to a neutral state,
stopping its viscous evolution. Instabilities are likely to clump
and destroy the disk eventually. As estimated by \citet{MPH01}
fall-back disks become neutralized at around  $10^4$ years. In our
models, the slower decrease in disk density and temperature leads
to neutrality at around $10^5$ years. CHN gave an explanation for
the upper cutoff of the spin period in terms of the transition
from a thin disk to an advection dominated accretion flow with
mass loss.

\section{DISCUSSION}

Our results show that for thin disks with bound-free opacities
prevailing most of the time and with a range of initial disk
masses ($M_d \sim 10^{-4}M_{\sun}$) and neutron star \emph{dipole}
magnetic moments ($\mu_{30}=1-10$) neutron stars do end up in the
period range $P=5-15$ s at ages $10^{4}-10^{5}$ years. The model
we introduced yields present day disks that are consistent with
the infrared observations for AXPs and SGRs (see, eg.
\citet{israel03,eikenberry01} and references therein), as we shall
detail in a subsequent paper. The observed luminosities require
that only a fraction of the mass flow rate is accreted onto the
neutron star in the current asymptotic phase.

The radial \citep{SS77} and temporal \citep{pringle91} properties of propeller
disks are different from the properties of disks in the mass-loss ``accretion'' regime.
The flow rate in the accretor solutions for the disk
decreases steeply both by losing mass due to accretion and by the
viscous spreading of the disk, whereas in the propeller solutions it is only the
``thinning'' of the disk by viscous spread that decreases the flow rate, while the disk mass remains constant.
It is possible that a fall-back disk loses mass also in the propeller regime because of the propeller
effect itself, depending on the fastness parameter. The simple power-law models for the
flow rate are working models which do not
take into account the coupling of the star to the structure of the disk. As shown
by \citet{FW} whether one obtains an AXP or not depends sensitively on the power-law index
$\alpha$. Extending the analysis to disk evolution in the propeller regime
provides favorable effective power-law indices $\alpha_p<1$.

We conclude that fall-back disks may well lead to the present day
properties of AXPs and SGRs, even in the available thin disk
models within the options of simple power law solutions. A more
realistic treatment would include effects of the magnetic field on
disk evolution, realistic opacities, including the iron line
opacity \citep{fryer} for the fall-back disk which is rich in
heavy elements, and the dependence of the accretion rate itself on
the conditions at the inner boundary of the disk, including the
effects of the motion of the inner radius \citep{spruit93}. All of
these effects would result in time evolution of the disk that
cannot be described by simple power laws. Taking account of the
complexities of the real problem is not likely to produce
qualitative disagreement with the evolutionary timescales and
tracks in the $P-\dot{P}$ diagram, as long as the torque is
efficient.

The fall-back disk models do not provide an explanation for the
AXP and SGR bursts or for the optical pulsations observed in
\objectname{4U 0142+614} \citep{KM02}. These phenomena are likely
to originate with processes in the crust or magnetosphere of a
neutron star with a strong magnetic field. We have shown that a
fall-back disk around a neutron star provides a viable mechanism
to get the observed period range at the AXP ages, provided that a
fraction $\eta \sim 10^{-2}$ of the mass flow is accreted.
Applying this model in conjunction with magnetar models for the
bursts and optical pulsations requires that the magnetar fields on
the surface of the neutron star are in higher multipoles. As we
have shown, the observed spindown rates and history of the AXPs
(or SGRs) are determined by disk torques rather than magnetic
dipole models. Such a hybrid model assumes that magnetospheric
generation of optical pulsations around a neutron star with a
surface dipole field of $\sim 10^{12}$ G and higher multipoles of
$\sim 10^{14}-10^{15}$ G is possible in the presence of the
fall-back disk. A demonstration of such a magnetospheric model is
beyond the scope of the present paper, and is to be attempted in
future work. We finally note that the extended quiet (low-noise)
spindown episodes of AXPs is compatible with the behavior of
accreting sources, in that the same AXPs have had high-level noise
characteristic of accreting sources; and there are well known
accreting sources like \objectname{4U 1907+09} which exhibit
similar extended quiet spindown \citep{baykal}.

\acknowledgments

We thank M.Hakan Erkut, \"{U}nal Ertan, Ersin G\"o\u{g}\"{u}\c{s}
and R.A.M.J. Wijers for useful discussions, the referee for helpful comments
and E.N. Ercan for her support and encouragement. This work was supported by
Bo\u{g}azi\c{c}i University Research Foundation under code 01B304 for KYE,
by  the Sabanc\i\ University Astrophysics and Space Forum,
by the High Energy Astrophysics Working Group of T\"{U}B\.{I}TAK
(The Scientific and Technical Research Council of Turkey)
and by the Turkish Academy of Sciences for MAA.

\clearpage

\clearpage

\end{document}